\documentclass{jfm}

\usepackage{graphicx}
\usepackage{natbib}
\usepackage{bbding}
\usepackage[caption=false]{subfig}
\ifCUPmtlplainloaded \else
  \checkfont{eurm10}
  \iffontfound
    \IfFileExists{upmath.sty}
      {\typeout{^^JFound AMS Euler Roman fonts on the system,
                   using the 'upmath' package.^^J}%
       \usepackage{upmath}}
      {\typeout{^^JFound AMS Euler Roman fonts on the system, but you
                   dont seem to have the}%
       \typeout{'upmath' package installed. JFM.cls can take advantage
                 of these fonts,^^Jif you use 'upmath' package.^^J}%
      }
  \else
  \fi
\fi

\ifCUPmtlplainloaded \else
  \checkfont{msam10}
  \iffontfound
    \IfFileExists{amssymb.sty}
      {\typeout{^^JFound AMS Symbol fonts on the system, using the
                'amssymb' package.^^J}%
       \usepackage{amssymb}%
       \let\le=\leqslant  
       \let\ge=\geqslant  \let\geq=\geqslant
      }{}
  \fi
\fi

\ifCUPmtlplainloaded \else
  \IfFileExists{amsbsy.sty}
    {\typeout{^^JFound the 'amsbsy' package on the system, using it.^^J}%
     \usepackage{amsbsy}}
    {}
\fi

\newsavebox{\astrutbox}
\sbox{\astrutbox}{\rule[-5pt]{0pt}{20pt}}

\newcommand\p{\ensuremath{\partial}}

\pubyear{2010}
\volume{650}
\pagerange{119--126}
\date{?; revised ?; accepted ?. - To be entered by editorial office}

\title[Aspect Ratio of Vortices in Rotating Stratified Flows]{The Universal Aspect Ratio of Vortices in Rotating Stratified Flows: Theory and Simulation}

\author[P. Hassanzadeh, P. S. Marcus, and P. Le Gal]%
{P\ls E\ls D\ls R\ls A\ls M\ns H\ls A\ls S\ls S\ls A\ls N\ls Z\ls A\ls D\ls E\ls H$^1$, \ns P\ls H\ls I\ls L\ls I\ls P\ns S.\ns M\ls A\ls R\ls C\ls U\ls S$^1$\thanks{Email address for correspondence: pmarcus@me.berkeley.edu} \break 
\and P\ls A\ls T\ls R\ls I\ls C\ls E\ns L\ls E\ns G\ls A\ls L$^2$}

\affiliation{$^1$Department of Mechanical Engineering, University of California, Berkeley, CA 94720, USA\\[\affilskip]
$^2$Institut de Recherche sur les Ph\'enom\`enes Hors Equilibre, UMR 7342, CNRS - Aix-Marseille Universit\'e, 49 rue F. Joliot Curie, 13384 Marseille, C\'edex 13, France}

\pubyear{????}
\volume{???}
\pagerange{????}
\date{?; revised ?; accepted ?. - To be entered by editorial office}
\begin{document}

\maketitle

\begin{abstract}
We derive a relationship for the vortex aspect ratio 
$\alpha$ (vertical half-thickness over horizontal length scale) for steady and slowly evolving vortices in rotating stratified fluids,
as a function of the Brunt-V\"ais\"al\"a frequencies within the vortex $N_c$ 
and in the background fluid outside the vortex $\bar{N}$, the Coriolis parameter $f$, and the Rossby number $Ro$ of the vortex: $\alpha^2 = Ro(1+Ro) \, f^2/(N_c^2-\bar{N}^2)$. 
This relation is valid for cyclones and anticyclones in either the cyclostrophic or geostrophic regimes;
it works with vortices in Boussinesq fluids or ideal gases, and the background density gradient need not be uniform.
Our relation for $\alpha$ has many consequences
for equilibrium vortices in rotating stratified flows. For example, cyclones must have $N_c^2 > \bar{N}^2$; weak anticyclones (with $|Ro|<1$) must have 
$N_c^2 < \bar{N}^2$; and strong anticyclones must have $N_c^2 > \bar{N}^2$. We verify our  relation for $\alpha$ with numerical  simulations of the
three-dimensional Boussinesq equations for a wide variety of vortices, including: 
vortices that are initially in (dissipationless) equilibrium and then
evolve due to an imposed weak viscous dissipation or density radiation;  
anticyclones created by the geostrophic adjustment of a patch of locally mixed density; cyclones created by fluid suction from a small localised region; vortices created from the remnants of the violent breakups of columnar vortices; and weakly non-axisymmetric vortices. 
The values of the aspect ratios of our numerically-computed vortices validate our relationship for $\alpha$, and generally they differ significantly
from the values obtained from the much-cited conjecture that $\alpha = f/\bar{N}$ in quasi-geostrophic vortices.
\end{abstract}

\begin{keywords}
\end{keywords}

\section{Introduction}\label{sec:intro}
Compact three-dimensional baroclinic vortices 
are abundant in geo- and astrophysical flows. 
Examples in planetary atmospheres include the rows of cyclones and anticyclones near Saturn's {\it Ribbon} \citep{Sayanagi10} and near $41^{\circ}$S on Jupiter  \citep{Humphreys07}, and 
Jupiter's anticyclonic Great Red Spot \citep{Marcus93}. In the  Atlantic Ocean 
meddies persist for years \citep{Armi88,McWilliams85}, and numerical simulations of the disks around protostars produce compact anticyclones \citep{Barranco05}. 
The physics that create, control, and decay these vortices is highly diverse, and the aspect ratios $\alpha \equiv H/L$ of these vortices range from flat ``pancakes" to nearly round (where $H$ is the vertical half-height and $L$ is the horizontal length scale of the vortex). However, we shall show that the aspect ratios of the vortices all obey a universal relationship. 


Our relation for $\alpha$ differs from previously published ones, including the often-used $\alpha = f/\bar{N}$, where $f$ is the Coriolis parameter, $N \equiv \sqrt{-\frac{g}{\rho} \frac{\p \rho}{\p z}}$ is the Brunt-V\"ais\"al\"a frequency, $g$ is the acceleration of gravity, $z$ is the vertical coordinate, $\rho$ is the density
for Boussinesq flows and potential density for 
compressible flows;
and a bar over a quantity indicates that it is the value of the unperturbed (i.e., with no vortices) background flow.
We shall show that  $\alpha = f/\bar{N}$ is not only incorrect by factors of 10 or more in some cases, but also that it is  misleading; 
it suggests that $\alpha$ depends only on the background flow and not on the properties of the vortex, so that {\it all} vortices embedded 
in the same flow (e.g., in the Atlantic or in the Jovian atmosphere) have the same $\alpha$. We shall show that this is {\it not} true. 
Knowledge of the correct relation for $\alpha$ is important. 
For example, there has been  debate over whether the color change, from white to red, of Jupiter's anticyclone Oval~BA, was due 
to a change in its  $H$ \citep{dePater10}. Measurements of the half-heights $H$ of planetary vortices are difficult, 
but $H$
can be accurately inferred from the correct relation for $\alpha$.
We validate our relation for $\alpha$ with 3D numerical simulations
of the Boussinesq equations. A companion paper by \citet{Aubert11} validates it with laboratory experiments and with observations of Atlantic ocean meddies and 
Jovian vortices. 
 
\section{Aspect Ratio: Derivation}\label{sec:scaling_law}
We assume that the rotation axis and gravity are parallel and anti-parallel to the vertical $z$ axis, respectively. 
We also assume that the vortices are in approximate cyclo-geostrophic balance horizontally and hydrostatic balance vertically (referred to hereafter as CG-H balance). Necessary approximations for CG-H balance are that the vertical $v_{z}$ and radial $v_{r}$
velocities are negligible compared to the azimuthal one $v_{\theta}$ (where the origin of the cylindrical coordinate system 
is at the vortex centre), that dissipation is  negligible, and that the flow is approximately steady in time. 
With these approximations, the radial $r$ and azimuthal $\theta$ components of Euler's equation in a rotating frame are 
\begin{eqnarray}
{{\partial p}/{\partial r}} = \rho v_{\theta} (f + v_\theta/r)  
\,\,\,\,\, {\rm and} \,\,\,\,\,\,\,
{{\partial p}/{\partial z}} = - \rho g,         
\label{rz1} 
\end{eqnarray}
where $p$ is the pressure. We have assumed that the vortex is axisymmetric, but will show later numerically that this approximation can be relaxed. Following the convention, we ignored the centrifugal term $\rho f^2 r / 4 \,\, \hat{\bf r}$ in equation~(\ref{rz1}) by assuming that the centrifugal buoyancy is much smaller than the gravitational buoyancy, i.e. that the rotational Froude number $f^2 d/(4g) \ll 1$, where $d$ is the characteristic distance of the vortex from the rotation axis \citep[see e.g.][]{Barcilon67}. The $\theta$-component of Euler's equation,  continuity equation, and the equation governing the dissipationless transport 
of (potential) density are all  satisfied by a steady, axisymmetric flow with $v_r = v_z =0$. As a consequence, 
equations~(\ref{rz1}) are the only equations that need to be satisfied for both Boussinesq   and  compressible flows. Thus,
our relation for $\alpha$ will also be valid for both of these flows. Far from the vortex, where $\mathbf{v}=0$, $p = \bar{p}$,  and $\rho=\bar{\rho}$, (\ref{rz1}) reduces to
\begin{eqnarray}
{{\partial \bar{p}}/{\partial r}} = 0 \,\,\,\,\, {\rm and}
\,\,\,\,\,\,\, {{\partial \bar{p}}/{\partial z}} = - \bar{\rho} g 
\label{rz2} 
\end{eqnarray}
showing that 
$\bar{p}$ and $\bar{\rho}$ are only functions of $z$. 
%
Subtracting equations~(\ref{rz2}) from (\ref{rz1}):
\begin{eqnarray}
{{\partial \tilde{p}}/{\partial r}} &=& \rho v_\theta (f + v_\theta/r) 
\label{rz3a} \\
{{\partial \tilde{p}}/{\partial z}} &=& - \tilde{\rho} g  
\label{rz3b} 
\end{eqnarray}
where $\tilde{p} \equiv  p - \bar{p}$ and $\tilde{\rho} \equiv \rho - \bar{\rho}$ are respectively the pressure and density anomalies. 
The centre of a vortex ($r=z=0$) is defined as the location on the $z$-axis where $\tilde{p}$ has its extremum,
so equation (\ref{rz3b}) shows that at the vortex centre
(denoted by a $c$ subscript) $\tilde{\rho}_c =0$ or ${\rho}_c = \bar{\rho}(0) \equiv \rho_o$. At the vortex boundary and outside  the vortex, where $v_{\theta}$ and $\tilde{\rho}$ are negligible, equations~(\ref{rz3a}) and (\ref{rz3b}) show that $\tilde{p} \simeq 0$. 

We define the pressure  anomaly's characteristic horizontal length scale (i.e. radius) as $L \equiv \sqrt{|4 \tilde{p}_c /(\nabla^2_{\perp} \tilde{p})_c |}$, 
where the subscript ${\perp}$ means {\it horizontal component}.  Integrating (\ref{rz3a}) from the vortex centre to its side boundary at $(r,z)=(L,0)$ approximately yields
\begin{eqnarray}
- \tilde{p}_c/L = \rho_o V_{\theta} (f+{V_{\theta}}/{R_v} )
\label{rz4a} 
\end{eqnarray}
where in the course of integration, $\rho$ has been replaced with $\rho_o$, which is exact for Boussinesq flows, and an 
approximation for fully compressible flows. Here $V_{\theta}$ is the characteristic peak azimuthal velocity, and $R_v$ is the approximate radius where the velocity has that peak. The analytical and numerically simulated vortices discussed below, meddies, and the laboratory vortices examined by \citet{Aubert11} all have $R_v =L$, but {\it hollow} vortices with quiescent interiors have $R_v \ne L$. For example, the Great Red Spot has $R_v \simeq 3 L$ \citep{Shetty10}. Similarly, integrating (\ref{rz3b}) from the vortex centre to its top boundary near $(r,z)=(0,H)$ approximately gives         
\begin{eqnarray}
- \tilde{p}_c/H = - g \tilde{\rho}(r=0, z=H), 
\label{rz4b}  
\end{eqnarray}
where $H$ is the pressure anomaly's characteristic vertical length scale (i.e. half-height), $H \equiv \sqrt{|2 \tilde{p}_c /(\p^2 \tilde{p} / \p z^2)_c|}$. 

Equations~(\ref{rz4a}) and (\ref{rz4b}) can be combined to eliminate $\tilde{p}_c$:
\begin{equation}
\frac{\rho_o V_{\theta} (f + V_{\theta}/R_v)}{H} = -\frac{g \tilde{\rho}(r=0, z=H)}{L}. 
\label{rz5}  
\end{equation}
Notice that this equation is basically the {\it thermal wind} equation, with the cyclostrophic term included (i.e. the gradient-wind equation \citep{Vallis06}), integrated over the vortex. Using the first term of a Taylor series, we approximate $\tilde{\rho}(r=0, z=H)$ on the right-hand side of (\ref{rz5}) with 
\begin{equation}
\tilde{\rho}(r=0, z=H)=\tilde{\rho}_c + H (\partial \tilde{\rho}/\partial z)_c = H [(\partial \rho/\partial z)_c - (\partial \bar{\rho}/\partial z)_c] = \rho_o H   (\bar{N}^2-N_c^2)/g
\label{rz6}
\end{equation}
where $\tilde{\rho} \equiv \rho - \bar{\rho}$ and $\tilde{\rho}_c=0$ have been  used. Note that in general, $\bar{N}(z)$ is a function of $z$; however, the only way in which $\bar{N}(z)$ is used in this derivation (or anywhere else in this paper) is at $z=0$ for evaluating $(\partial \bar{\rho}/\partial z)_c$. Therefore, rather than using
the cumbersome notation $\bar{N}_c$, we simply use $\bar{N}$.

Using (\ref{rz6}) in equation~(\ref{rz5}) gives our relation for $\alpha$:
\begin{eqnarray}
\alpha^2 \equiv \left (\frac{H}{L} \right )^2 = \frac{Ro \, \left [1 + Ro \,
 (L/R_v) \right ]}{N_c^2-\bar{N}^2} f^2 
\label{scale}
\end{eqnarray}
where the Rossby number defined as $Ro \equiv V_{\theta}/(f L)$ can be well approximated as $Ro = \omega_c/(2f)$, $\omega_c$ being the vertical component of vorticity at the vortex centre. Defining the Burger number as $Bu \equiv (\bar{N} H/(fL))^2$, equation~(\ref{scale}) may be as well rewritten as
\begin{eqnarray}
[(N_c/\bar{N})^2-1] Bu =Ro \,(1+(L/R_v)Ro) 
\end{eqnarray}

Equation~(\ref{scale}) shows that $\alpha$ depends on two properties of the vortex: $Ro$ and the {\it difference} between the Brunt-V\"ais\"al\"a frequencies inside the vortex (i.e. $N_c^2$) and outside the vortex (i.e. $\bar{N}^2$).   
Note that to derive relation~(\ref{scale}), no assumption has been made on the compressibility of the flow, Rossby number smallness, dependence of $\bar{N}$ on $z$, or the magnitude of $N_c/\bar{N}$. 
Therefore, equation~(\ref{scale}) is applicable to 
Boussinesq, anelastic \citep{Vallis06}, and fully compressible flows, 
cyclones (i.e., $Ro>0$) and anticyclones (i.e., $Ro <0$), and geostrophic and cyclostrophic flows.     
In the cyclostrophic limit (i.e., $|Ro| \gg 1$) with $N_c=0$ and $R_V=L$, $Ro \, (1+ Ro) \rightarrow Ro^2$, hence equation~(\ref{scale}) 
becomes $V_{\theta} = H \bar{N}$, agreeing with the findings of \citet{Billant01} and others. 
Equation~(\ref{scale}) is easily modified for use with discrete layers of fluid rather
than a continuous stratification, and in that case agrees with the theoretical work of \citet{Nof81} and \citet{Carton01}.   

Equation~(\ref{scale}) has several consequences for equilibrium vortices.
For example, because the right-hand side of (\ref{scale}) must be positive, cyclones  must have $N_c^2 \ge \bar{N}^2$. Another consequence is
that anticyclones with   $-Ro < R_v/L$, must have $N_c^2 \le \bar{N}^2$, and anticyclones with $- Ro > R_v/L$, have $N_c^2 \ge \bar{N}^2$. 
In addition, equation~(\ref{scale}) is useful for astrophysical and geophysical observations of vortices in which some of the vortex properties are difficult to measure.
For example, $N_c$ is difficult to measure in some ocean vortices \citep{Aubert11}, and $H$ is difficult to determine in some satellite observations of atmospheric vortices \citep{dePater10}, but their values
can be inferred from equation~(\ref{scale}).

Note that $N_c$ is a measure of the mixing within the vortex; if the density is not mixed with respect to the background flow, then $N_c \rightarrow \bar{N}$ (and the vortex is a tall, barotropic Taylor column); 
if the density is well-mixed within
the vortex so the (potential) density is uniform inside the vortex, then $N_c \rightarrow 0$ (as in the experiments of \citet{Aubert11}); if $N_c^2 > \bar{N}^2$ (as
required by cyclones), then the vortex is more stratified than the background flow.       

\section{Previously Proposed Scaling Laws}\label{sec:previous}
Other relations for $\alpha$ that differ from our equation~(\ref{scale}) have been published previously, and the most frequently cited one is $\alpha \equiv H/L=f/\bar{N}$. This relationship is inferred  from Charney's equation for the quasi-geostrophic (QG) potential vorticity \citep[equation~(8) in][]{Charney71} that was derived for flows with  $|Ro| \ll 1$ and $N_c/\bar{N} \simeq 1$.
Separately re-scaling the vertical and horizontal coordinates of the potential vorticity equation, and then assuming that the the vortices are isotropic in the
re-scaled (but not physical) coordinates, one obtains
the alternative scaling $\alpha  = f/\bar{N}$.   
Numerical simulations of the QG equation for some initial conditions have produced turbulent vortices with 
$H/L \approx f/\bar{N}$ \citep[c.f.,][]{McWilliams99,Dritschel99,Reinaud03}, even though significant anisotropy in the re-scaled coordinates was observed in 
similar simulations \citep{McWilliams94}. The constraints under which the QG equation is derived 
are very restrictive; for example, none of meddies or laboratory vortices studied by \citet{Aubert11} meet these requirements 
because $N_c/\bar{N}$ is far from unity. Therefore, it is not surprising that none of these vortices, including the laboratory vortices, 
agree with $\alpha \approx f/\bar{N}$, but instead have $\alpha$ in accord with relation~(\ref{scale}) \citep{Aubert11}.  

The constraints under which our equation~(\ref{scale}) for $\alpha$ is derived are far less restrictive 
than those used in  deriving Charney's QG equation (and we never need to assume isotropy). In particular, one of {\it several constraints} needed for deriving 
Charney's QG equation is the scaling required for the potential temperature (his equation~(3)), which written in terms of the potential density is 
\begin{equation}
\tilde{\rho}/\bar{\rho} = - (f/g) (\partial \psi/\partial z), \label{all}
\end{equation} 
where $\psi$ is the stream function of horizontal velocity. This  constraint {\it alone} (which is effectively the thermal wind equation) implies our 
relationship~(\ref{scale}) for $\alpha$. To see this, in equation~(\ref{all}) replace $\psi$ with $V_{\theta} L$, 
$\partial /\partial z$ with $1/H$,  $V_{\theta}$ with $Ro f L$, and  $\tilde{\rho}/\bar{\rho}$ with 
$H \{(\partial \rho/\partial z)_c -  (\partial \bar{\rho}/\partial z)\}/\bar{\rho} = (H/g) (\bar{N}^2 - N_c^2)$.  
With these replacements, equation~(\ref{all}) immediately gives 
\begin{eqnarray}
(H/L)^2 = Ro \, f^2 /(N_c^2 -\bar{N}^2), 
\end{eqnarray}
which is the small $Ro$ limit of equation~(\ref{scale}). 

\citet{Gill81} also proposed a relationship for $\alpha$ that differs from ours. He based his relation for $\alpha$ on a model  2D zonal flow (that is, not  an axisymmetric vortex, but rather
a 2D vortex) and found that  ${\alpha}$ was proportional to $Ro \,  f/\bar{N}$. To determine $\alpha$, Gill derived separate solutions for the flow inside and outside 
his 2D model vortex, which he assumed was dissipationless and  in geostrophic and hydrostatic balance. Despite the fact that Gill's {\it published} relation for $\alpha$, obtained from the {\it outside} solution, differs from ours, we can show
that his solution for the flow {\it inside} his 2D vortex satisfies {\it our} scaling relation for $\alpha$. Gill's solution for the zonal 
velocity (which is in the $y$ direction)
is $v=-(f/a) x$ (his equation~(5.14) in dimensional form). His density anomaly is $\tilde{\rho} = \rho_o \, (\bar{N}^2/g) \, z$ (i.e. 
within the 2D vortex, $\rho=\rho_o$). The equation for $v$ gives $\omega_z=-(f/a)$, and therefore $Ro \equiv \omega_c /(2f)=-1/(2a)$. Substituting
$v$ and $\tilde{\rho}$ into the equations for geostrophic and hydrostatic balance, gives $\p \tilde{p}/\p x = - \rho_o (f^2/a) \, x$ and $\p \tilde{p}/\p z = -\rho_o \, \bar{N}^2$, respectively. Using the definitions of $H$ and $L$ from section~\ref{sec:scaling_law} along with $Ro=-1/(2a)$, we obtain $\alpha^2 = - Ro \, f^2/\bar{N}^2$, 
which is our relation~(\ref{scale}) in the limit of small $Ro$, $L=R_v$, and $N_c=0$ (which are the constraints under which Gill's solution is obtained).
Gill's scaling for $\alpha$ is derived from the flow {\it outside} the vortex, which he derived by requiring that both the                   
tangential velocity and 
density are continuous at the interface between the inside and outside solutions. 
In general, this {\it over-constrains} the dissipationless flow (which only requires pressure and normal velocity to be continuous) -- see for example the vortex solution in  \citet{Aubert11} in which the pressure and normal 
component of the velocity are continuous at the interface, but not the density or tangential velocity.
The extra constraints force the solution outside Gill's vortex 
to have additional, (unphysical) length scales, resulting in Gill's relation for $\alpha$ differing from ours. \citet{Aubert11} show that Gill's relationship for $\alpha$
does not fit their laboratory experiments, meddies, or Jovian vortices.
We examine the accuracy of both Charney's and Gill's relationships in section~\ref{sec:results}.
 
\section{Gaussian Solution to the Dissipationless Boussinesq Equations}
\label{sec:analytic}
It is easy to find closed-form solutions to the steady, axisymmetric, dissipationless Boussinesq equations (e.g. \citet{Aubert11}).
One solution that we shall use to generate initial conditions for our initial-value codes is the {\it Gaussian} vortex with
$\tilde{p} = \tilde{p}_c  \, \exp[-(z/H)^2 -(r/L)^2]$  and $v_r = v_z =0$  
(where $\tilde{p}_c$, $H$, and $L$ are arbitrary constants). 
Then, $\tilde{\rho}$ is found from
$\partial \tilde{p}/\partial z$ using equation~(\ref{rz3b}), and $v_{\theta}$ is found from 
$\partial \tilde{p}/\partial r$ using equation~(\ref{rz3a}) with 
$\rho$ replaced by  $\rho_o$.
This Gaussian vortex exactly obeys our relationship~(\ref{scale}) for $\alpha$ when
the Rossby number is defined as before as $Ro \equiv \omega_c/(2f)$, when $R_v$ is set equal to $L$, and when the 
vertical and horizontal scales  are 
defined as in section~\ref{sec:scaling_law}.
Note that $N(r,z)$ within the vortex  is not uniform,
that $N_c^2 = \bar{N}^2 - 2 \tilde{p}_c/(\rho_o H^2)$, and that the vortex is {\it shielded}.
By {\it shielded}, we mean that there is a ring of cyclonic (anticyclonic) vorticity around the anticyclonic (cyclonic) core in each horizontal plane, and therefore at each $z$, circulation due to the vertical component of the vorticity is zero (i.e. the vortices are {\it isolated}).     
The Gaussian vortex could be a cyclone or an anticyclone depending on the choice of constants.
This vortex is well-studied and has been widely used to model isolated vortices, especially in the oceans \citep[e.g.][]{Gent86,Morel97,Stuart11}.

\section{Numerical Simulation of the Boussinesq Equations}\label{sec:num_sim}
We have used 3D numerical simulations to verify our relation~(\ref{scale}) for $\alpha$ in a  Boussinesq flow with constant $\bar{N}$
and $f$.
We include dissipation and solve the equations in a rotating-frame in Cartesian coordinates \citep{Vallis06}:
\begin{equation}
\nabla \cdotp \mathbf{v} = 0, \; \; \; \; \; \frac{D \mathbf{v}}{D t} = -\frac{\nabla p}{\rho_o} + \mathbf{v} \times f \mathbf{\hat{z}} - \frac{\tilde{\rho}}{\rho_o} g \mathbf{\hat{z}} + \nu \nabla^2 \mathbf{v}, \; \; \; \; \; \frac{D \tilde{\rho}}{D t} = \rho_o \, \frac{\bar{N}^2}{g} w - \frac{\tilde{\rho}}{\tau_{rad}},  
\label{eq1}
\end{equation} 
where $D/Dt = \p / \p t + \mathbf{v} \cdot \nabla$, and $\mathbf{v} = (u,v,w)$ (Notice that throughout this paper, we use $v_z$ and $w$ for the vertical component of velocity in the cylindrical and Cartesian coordinates, respectively.) We include kinematic viscosity $\nu$, 
but neglect the diffusion of density because diffusion is slow (e.g., for salt-water the Schmidt number is $\sim 700$). 
Instead, inspired by astrophysical vortices (e.g., Jovian vortices or vortices of protoplanetary disks) for which 
thermal radiation is the main dissipating mechanism, we have added the damping 
term $-\tilde{\rho}/\tau_{rad}$ to the density equation to model {\it radiative dissipation} where $\tau_{rad}$ is radiative dissipation time scale. 

A pseudo-spectral method with $256^3$ modes is used to solve equations~(\ref{eq1}) in a triply periodic domain (which was chosen to be $10$ to $20$ times larger than the vortex in each direction). Details of the numerical method is the same as \citet{Barranco06}. The results of our triply periodic code are qualitatively, and in most cases quantitatively, the same as solutions we obtained with a code with no-slip vertical boundary conditions. That is because our vortices are far from the vertical boundaries, and therefore the Ekman circulation is absent.  


\section{Numerical Results for Vortex Aspect Ratios}
As shown in table \ref{tab}, we have examined the aspect ratios of vortices in four types of initial-value numerical experiments. The goal of these simulations is to determine how well the aspect ratios $\alpha$ of vortices obey our relation~(\ref{scale}) as they evolve in time. 

\begin{table}
\begin{center}
\def~{\hphantom{0}}
	\begin{tabular}{ccccc | cccccc}
   Case & $\bar{N}/N_c$ & $H /L$                   & $Ro$    & $E_k(10^{-5})$ & 
   Case & $\bar{N}/N_c$ & $H/L$     & $\tau_{rad}$ & $Ro$    & $E_k(10^{-5})$   \\ [2pt]  \hline 
A1   & $1/0        $ & $16/ 8$   				   & $-0.2$  & $25 $            & 
A18  & $3.33/3.17    $ &$1.44/0.72$& $\infty$     & $-0.22$ & $4 $              \\ 
A2   & $1/0        $ & $16/ 8$   					& $-0.2$  & $12.5 $            & 
A19  & $10/10.5  $ &$0.16/0.08$& $\infty$     & $+0.87$  & $250$              \\
A3   & $1/0        $ & $16/ 8$   & 					$-0.2$  & $6.25 $            &
A20  & $3.33/3.49  $ &$1.44/0.72$& $1656$       & $+0.14$ & $0   $              \\ 
A4   & $1/0.5      $ & $16/ 8$   					& $-0.14$  & $25 $            & 
A21  & $1/1.0475   $ &$16/8$     & $1656$       & $+0.015$ & $0   $              \\
A5   & $1/0.5      $ & $4 / 2$   					& $-0.14$  & $25 $            &
A22  & $3.33/3.17  $ &$1.44/0.72$& $1237$     & $-0.22$  & $0   $              \\ 
A6   & $20/19.99 $ &$0.05/0.02$& 					$-0.11$  & $500$            & 
A23  & $3.33/3.49  $ &$1.44/0.72$& $120$        & $+0.14$  & $0   $              \\
A7   & $1/0        $ &$20/12$    			& $-0.12$  & $11.1 $            &
A24  & $3.33/3.17  $ &$1.44/0.72$& $120$        & $-0.22$  & $0   $              \\
A8   & $1/0      $ &$16/8 $      			& $-0.2$  & $25$            &
B1   & $1/0        $ &$20/12$    & $\infty$     & $-0.12$  & $4 $              \\
A9   & $1/0      $ &$12/12$    			   & $-0.04$  & $11.1$            &
B2   & $1/0        $ &$20/12$    & $\infty$     & $-0.12$  & $4 $              \\
A10  & $20/19.9 $ &$0.03/0.03$					& $-0.2$  & $222.2$           &
B3   & $1/0        $ &$20/12$    & $\infty$     & $-0.12$  & $0 $              \\
A11  & $20/19.9 $ &$0.03/0.03$					& $-0.2$  & $11.1$            &
C1   & $1/0.761   $ &$6.09/5.67$& $\infty$     & $+0.33$& $49.7 $              \\
A12  & $3.53/2.5   $ &$0.96/0.96$					& $-0.5$  & $39.3$            &
C2   & $1/0.709    $ &$5.91/5.67$    & $\infty$     & $+0.35$  & $45.9$              \\
A13  & $3.54/3.45  $ &$0.96/0.96$					& $-0.023$ & $39.3$            &
D1   & $2.5/2.5    $ &$40/2$     & $\infty$     & $-0.75$  & $25.4  $              \\
A14  & $1.56/1.64  $ &$6.55/3.28$					& $+0.0386$& $3.91$            &
D2   & $2.5/2.5    $ &$40/2$     & $\infty$     & $-0.75$  & $12.7$              \\
A15  & $0.5/0.75   $ &$16/8$     				& $+0.0477$& $4 $            &
D3   & $1.67/1.67    $ &$90/4.5$   & $\infty$     & $-0.5$   & $16.9$              \\
A16  & $0.5/0.55   $ &$16/8$     					& $+0.0083$& $4 $            &
D4   & $5/5     $ &$10/0.5$   & $\infty$     & $-1$     & $50.7$              \\
A17  & $3.33/3.49  $ &$1.44/0.72$					& $+0.15$  & $4 $            &
D5   & $20/20   $ &$0.6/.03$    & $\infty$     & $+0.8$  & $24.4 $              \\
   \end{tabular}
	\caption{Parameters of the background flows and of the vortices at the ``initial" time. For Cases A, B, and D the ``initial" time is $t=0$, and for Case C the ``initial" time is $t=t_{off}$. All values are in CGS units. For all cases, $f=5$ rad/s, $g=980 \, \, \rm{m/s}^2$, and $\rho_o=1 \, \, \rm{g/cm}^3$. Ekman number is defined as $E_k \equiv \nu/(f L^2)$. See text for the difference between Cases B1 and B2. For Case C1, $t_{off}=60$ s and $Q=-64 \, \, \rm{cm^3/s}$, and for Case C2, $t_{off}=30$ s and $Q=-128 \, \, \rm{cm^3/s}$.}
   \label{tab}
	\end{center}
\end{table}

\subsection{Case A: Run-Down Experiments}
In this case, our initial
condition is the velocity and density anomaly of the Gaussian vortex from section~\ref{sec:analytic} that is an exact equilibrium of the dissipationless Boussinesq equations with constant $f$ and $\bar{N}$. These are ``run-down'' experiments because they are carried out {\it either} with radiative dissipation (i.e., finite $\tau_{rad}$) {\it or} viscosity, but not both. Due to the weak dissipation, the vortices slowly evolve (decay) and do not remain Gaussian. Also, as a result of the dissipation (and decay), a weak secondary flow is induced (i.e. non-zero $v_r$ and $v_z$).

\subsection{Case B: Vortices Generated by Geostrophic Adjustment}
This case is motivated by vortices produced from the geostrophic adjustment of a locally mixed patch of density, e.g. generated from diapycnal mixing \citep[see e.g.][]{McWilliams88,Stuart11}. Our flow is initialised with $\mathbf{v}=0$ and $\tilde{\rho} \neq 0$. For Cases B1 and B3 the initial $\tilde{\rho}$ 
is that of the Gaussian vortex discussed in
section~\ref{sec:analytic}. But here, the initial flow is far from equilibrium because $\mathbf{v} \equiv 0$. In Case B2, the initial $\tilde{\rho}$ is Gaussian in $r$, but has a top-hat function in $z$ (for this case, the
initial $H$ is defined as the half-height of the top-hat function). It is observed in the numerical simulations that geostrophic adjustment quickly produces shielded vortices.
 
\subsection{Case C: Cyclones Produced by Suction}
Injection of fluid into a rotating flow generates anticyclones \citep{Aubert11}, while suction produces cyclones. 
We simulate suction by modifying the continuity equation in (\ref{eq1}) 
as $\nabla \cdot \mathbf{v} = Q(\mathbf{x},t)$ 
where $Q$ is a specified suction rate function. 
The flow is initialised with $\mathbf{v}=\tilde{\rho}=0$. Suction starts at $t=0$ over a spherical region with radius of $6 \, \rm{cm}$ and is turned off at time $t_{off}$. A shielded cyclone is produced and strengthened during the suction process. As mentioned at the
end of section~\ref{sec:scaling_law}, for $Ro>0$, relation~(\ref{scale}) requires $N_c>\bar{N}$, which we have shown in the numerical simulations that the initial suction creates. 
Cases C1 and C2 have different suction rates and $t_{off}$, but the same total sucked volume of fluid, and it is observed that the produced cyclones are similar. 

\subsection{Case D: Vortices Produced from the Breakup of Tall Barotropic Vortices}
The violent breakup of tall barotropic ($z$-independent) vortices in rotating, stratified flows can produce stable compact vortices \citep[see e.g.][]{Smyth98}. 
In Case D, our  flows are initialised with an unstable 2D columnar vortex with $v_{\theta}=Ro \, f \, r \exp(-(r/L)^2)$ and $\tilde{\rho}=0$ (for this case, the initial $H$ is the vertical height of the computational domain). Note that the initial columnar vortex is shielded. Noise is added to the initial velocity field to hasten instabilities. 
The vortex breaks up and then the remnants equilibrate to one or more compact shielded vortices (in each case, only the vortex with the largest $|Ro|$ is considered in section~\ref{sec:results}).       


\section{Aspect Ratio: Numerical Simulations}\label{sec:results}
In all cases, vortices reach quasi-equilibrium and then slowly decay due to viscous or radiative dissipation except for Case B3 which is dissipationless and evolves only due to geostrophic adjustment. As a result, $Ro$ decreases, and the mixing of density in the vortex interior changes (i.e., $N_c$ changes). Therefore, it is not surprising that the aspect ratio $\alpha$ also changes in time. Quasi-equilibrium is reached in Case A almost immediately. In Case B, vortices quickly form and come to quasi-equilibrium after geostrophic adjustment. Quasi-equilibrium is achieved following the geostrophic and hydrostatic adjustments after $t_{off}$ in Case C, and (much longer) after the initial instabilities in Case D.      

For each case, we use the results of the numerical simulations to calculate
$Ro(t) \equiv \omega_c(t)/(2f)$ and $N_c(t) \equiv \sqrt{\bar{N}^2-g(\p \tilde{\rho}(t)/\p z)_c/{\rho}_o}$. We compute $L(t)$ and $H(t)$ from the numerical solutions 
using their definitions given in section~\ref{sec:scaling_law}.
Calculating $L$ based on $\nabla_{\perp}^2$ rather than just $r$-derivatives is useful for non-axisymmetric vortices. For example,
due to a small non-axisymmetric perturbation added to the initial condition of  Case A8, the vortex went unstable and produced a tripole \citep{vanHeijst89}. Cases C1 and C2 also produced non-axisymmetric vortices. We define the {\it numerical} aspect ratio as $\alpha_{\rm NUM}(t) \equiv H(t)/L(t)$.
We define the {\it theoretical} aspect ratio $\alpha_{\rm THR}$ from equation~(\ref{scale}) using $Ro(t)$ and $N_c(t)$ extracted from the numerical results 
and the (constant) values of $f$ and $\bar{N}$.   

\begin{figure}
  \centerline{\includegraphics[width=1.\textwidth]{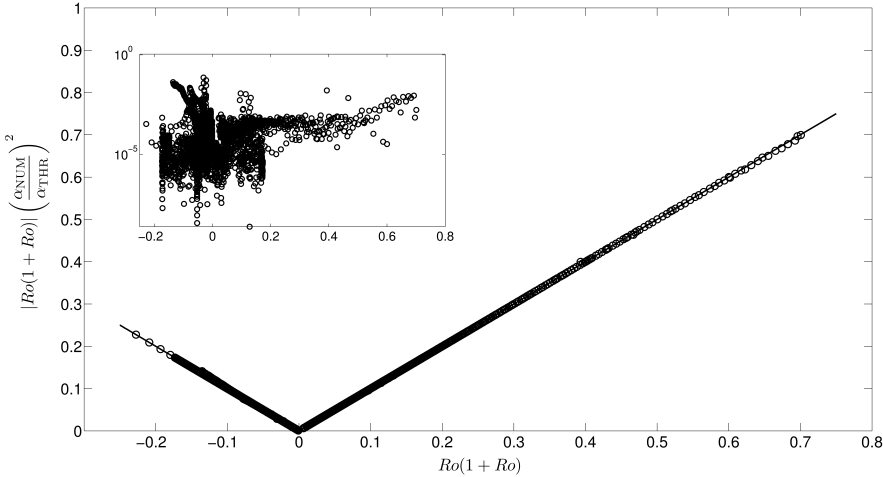}}
  \caption{Comparison of $\alpha_{\rm NUM}$ with $\alpha_{\rm THR}$ (see text for definitions). The circles show the value of $|Ro(1+Ro)|(\alpha_{\rm NUM}/\alpha_{\rm THR})^2$ and the straight lines show the value of this expression if $\alpha_{\rm NUM} \equiv \alpha_{\rm THR}$. All $4122$ data points (circles) collapse on the straight lines (and densely cover them), validating our equation~(\ref{scale}). Data points are recorded one inertial period ($=4\pi/f$) after the initial time (as defined in table~\ref{tab}) in Cases A-C, and $50$ inertial periods after $t=0$ in Case D. Note that all of our simulated vortices have $L=R_v$. The horizontal axis in the inset 
is the same as in the main figure; the inset's vertical axis is the relative difference $|1-(\alpha_{\rm NUM}/\alpha_{\rm THR})^2|$ (which is $< 0.07$).
({\it n.b.}, the left-most plotted point has $Ro(1+Ro) \simeq -0.25$ due to the mathematical tautology that $Ro(1+Ro)\geq -0.25$ for all values of $Ro$.)}
\label{fig:ar}
\end{figure}

Figure \ref{fig:ar} shows how well $\alpha_{\rm THR}$ agrees with $\alpha_{\rm NUM}$. The inset in figure \ref{fig:ar} shows that the relative difference between the two values, calculated as $|1-(\alpha_{\rm NUM}/\alpha_{\rm THR})^2|$, is smaller than $0.07$. For each case, the maximum difference occurs at early times or during instabilities.
 
\begin{figure}
  \centering
  \subfloat[]{\label{fig3a}\includegraphics[width=0.5\textwidth]{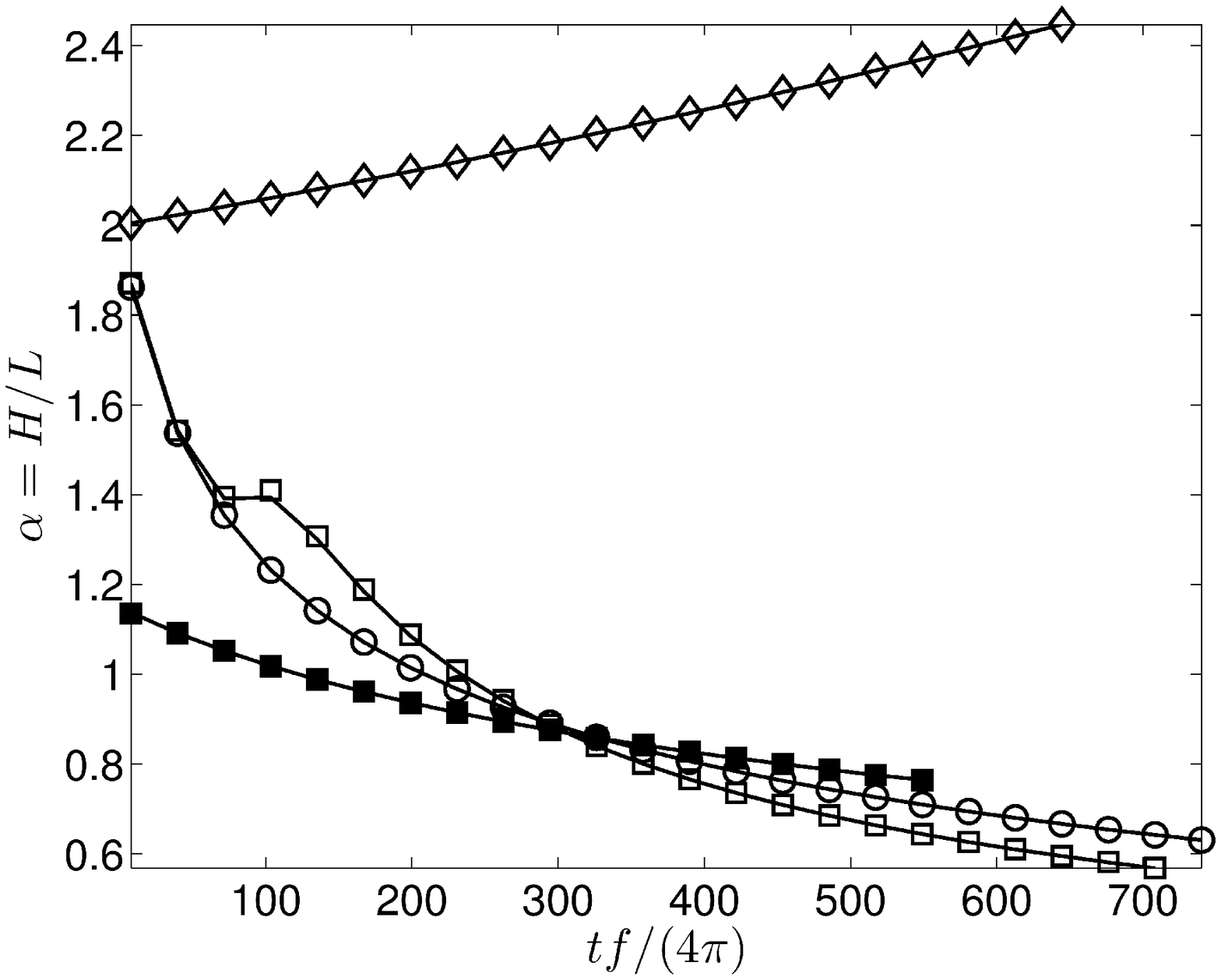}}                
  \subfloat[]{\label{fig3b}\includegraphics[width=0.5\textwidth]{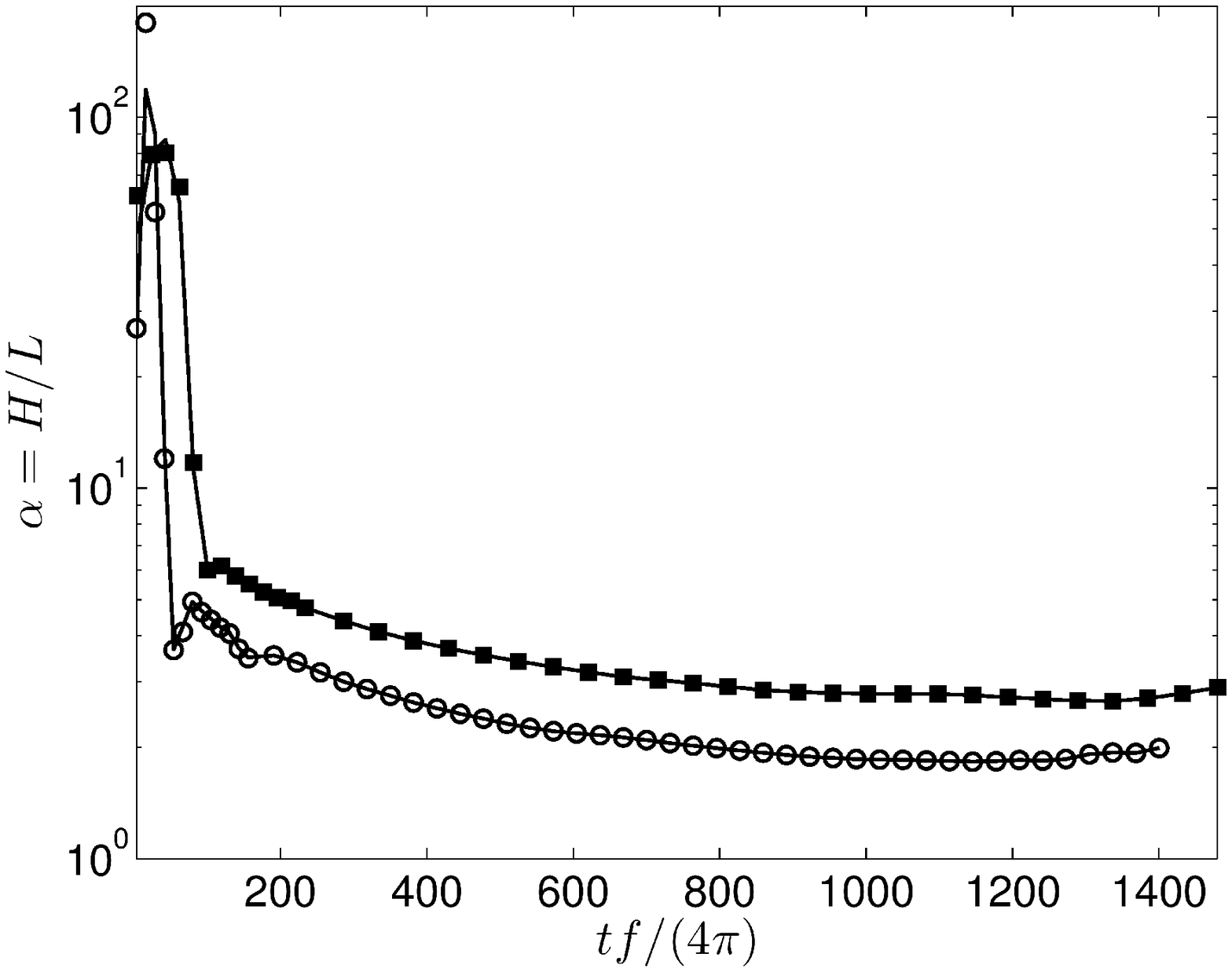}} 
  \caption{Time evolution of $\alpha_{\rm THR}$ (continuous curves), and $\alpha_{\rm NUM}$ for Cases (a) A1 $\circ$, A8 $\square$, B1 $\blacksquare$, and A20 $\diamond$, (b) D1 $\circ$, and D3 $\blacksquare$. Cases A1, A8 and B1 have $f/\bar{N}=5$ and Case A20 has $f/\bar{N}=1.5$ which differ significantly from $\alpha$. Case D1 has $f/\bar{N}=2$ and Case D3 has $f/\bar{N}=3$ which agree with $\alpha$ only at the late times.}     
  \label{fig:cmp}
\end{figure}

Figure~\ref{fig:cmp} compares the values of $\alpha_{\rm NUM}$ with $\alpha_{\rm THR}$ as a function of time for six cases.
The figure starts at time $t=0$, so it includes vortices which are not in CG-H equilibrium to   
highlight the situations for which relationship~(\ref{scale}) for $\alpha$ is {\it not} good due to violation of its assumptions. 
Cases A1, B1, and A20 in figure~\ref{fig3a} exhibit excellent agreement with our theoretical prediction for $\alpha$, while Case A8 
shows a small deviation starting around $t=80(4\pi/f)$. This deviation is a result of the vortex going unstable at this time (accompanying by relatively large $v_r$ and $v_z$) and forming a tripolar vortex .
After the tripole comes to CG-H equilibrium, its $\alpha$ once again agrees with theory. As the vortices dissipate, and $Ro$ and $N_c$ change, $\alpha$ can either decrease in time (c.f., Case A1) or increase (c.f., Case A20).

Figure \ref{fig3b} shows Cases D1 and D3 from time $t=0$. The remnant vortices that formed from the violent break-up of the columnar vortices are initially far from the CG-H balance. As a result, the value of $\alpha_{\rm THR}$ at these early times does not fit well with the values of $\alpha_{\rm NUM}$. However, after the CG-H balance is established in the remnants, our theoretical relationship~(\ref{scale}) becomes valid and $\alpha_{\rm THR}$ agrees well with $\alpha_{\rm NUM}$.              

Figure~\ref{fig3a} shows that the alternative scaling relation based on Charney's QG equation, $\alpha = f/\bar{N}$, is not a good fit
to our numerical data. Cases A1, A8, and B1 all have $=5$ which is obviously far from the measured aspect ratio of these vortices. Case A20 has $f/\bar{N}=1.5$ which again does not agree with $\alpha_{\rm NUM}$. In fact, in all four cases, the difference between $\alpha_{\rm NUM}(t)$ and $f/\bar{N}$ increases by time, while $\alpha_{\rm THR}(t)$ always remains close to $\alpha_{\rm NUM}(t)$. For other cases in table~\ref{tab}, it has been observed that for vortices which are in CG-H equilibrium, $\alpha_{\rm NUM}/(f/\bar{N})$ can be as large as $9.56$ and as small as $0.11$. The data displayed in figure~\ref{fig3b} were carefully ``cherry-picked'' from all of our runs because
they are unusual in that  $\alpha \rightarrow f/\bar{N}$ after a long time. The fluid within the remnants strongly mixed with the background fluid, 
so at late times $N_c \rightarrow \bar{N}$ and $Ro$ significantly decreases and therefore the conditions needed for the validity of Charney's QG equation are approached. 
Whether these results are a fluke and whether $\alpha \rightarrow f/\bar{N}$ for all  vortices that are created by {\it one particular method} is not yet clear.
The physics governing these vortices is currently be investigated and will be discussed in a future paper.   

Gill's model \citep{Gill81}, discussed in section~\ref{sec:previous}, is not a good fit to any of our numerically computed vortices. For example, the value of $\alpha_{\rm NUM}(t)/(Ro(t) f/\bar{N})$ is between $2$ and $8$ for Case A1; $2$ and $9$ for Case A8; $80$ and $160$ for Case A20; and $4$ and $7$ for Case B1. The much larger error observed for Case A20 is due to the fact that unlike the other three cases, $N_c$ is far from $0$ in this case, and Gill's derivation does not incorporate $N_c \neq 0$.                

\section{Conclusion}
We have derived a new relationship~(\ref{scale}) for the aspect ratio $\alpha$ of baroclinic vortices in cyclo-geostrophic and hydrostatic (CG-H) equilibrium and used numerical initial-value simulations of the Boussinesq equations to validate this relation for a wide variety of unforced quasi-steady vortices generated and dissipated with different mechanisms. Our new relationship shows that $\alpha$ depends on the background flow's Coriolis parameter $f$ and Brunt-V\"ais\"al\"a frequency  $\bar{N}$, as well as properties of the vortex, including $Ro$ and  $N_c$. Thus, it shows that all vortices embedded in the same {\it background} flow do {\it not} have the same aspect ratios. In a companion paper, \citet{Aubert11} verify the new relationship with laboratory experiments and show it to be consistent with observations of Atlantic meddies and Jovian vortices.

Equation~(\ref{scale}) for $\alpha$ has several consequences. For example, it shows that for cyclones ($Ro>0$), $N_c$ must be greater than $\bar{N}$, that is, the fluid within a cyclone must
be {\it super-stratified} with respect to the background stratification. Mixing usually de-stratifies the flow over a local region, and therefore cannot produce cyclones. This may explain why there are more anticyclones than cyclones observed in nature. We numerically simulated local suction to create cyclones, and we found that suction creates a large envelope of super-stratified flow around the location of the suction and when the suction is stopped, the CG-H adjustment makes cyclones. Details of these simulations and results of an ongoing laboratory experiment will be presented in subsequent publications.   

It is widely quoted that vortices obey the quasi-geostrophic scaling law $\alpha = f/\bar{N}$ (i.e. Burger number $Bu=1$). This is inconsistent with our relationship which written in terms of $Bu$ is $Bu=Ro(1+Ro)/[(N_c/\bar{N})^2-1]$. We found that, with the exception of one family of vortices, the quasi-geostrophic scaling law was not obeyed by the vortices studied here (and by \citet{Aubert11}), and could be incorrect by more than a factor of $10$. Another relationship proposed by \citet{Gill81} was also found to produce very poor predictions of aspect ratio.    

We found that $\alpha$ can either increase or decrease as the vortex decays, and our relationship~(\ref{scale}) shows that the dependence of $\alpha$ on $N_c$ is specially sensitive when $N_c$ is at the order of $\bar{N}$, as it is for meddies and Jovian vortices \citep{Aubert11}. Our simulations showed that $N_c$ was determined by the secondary circulations within a vortex and that those circulations are controlled by the dissipation. In a future paper we shall report on the details of how dissipation determines the secondary flows and the temporal evolution of $N_c$, both of which are important in planetary atmospheres, oceanic vortices, accretion disk flows, and planet formation \citep{Barranco05}. \\  

This work used an allocation of computer resources from the Extreme Science and Engineering Discovery Environment (XSEDE), which is supported by National Science Foundation 
grant number OCI-1053575. We acknowledge support 
from the NSF 
AST and ATI Programs, and from the NASA Planetary Atmospheres Program.
P. H. was supported in part by the 
Natural Sciences and Engineering Research Council of Canada through a PGS-D scholarship. P.S.M. thanks the France-Berkeley Fund and Ecole Centrale Marseille. P.L.G. thanks the 
Russel Severance Springer Professorship endowment and the Planetology National Program (INSU, CNRS).


\bibliographystyle{jfm}
\bibliography{pbib}

\end{document}